# Comparison of the effects of investor attention using search volume data before and after mobile device popularization

Jonghyeon Min

Abstract

  In this study, we will study investor attention measurement using the Search Volume Index in the recent market. Since 2009, the popularity of mobile devices and the spread of the Internet have made the speed of information delivery faster and the investment information retrieval data for obtaining investment information has increased dramatically. In these circumstances, investor attention measurement using search volume data can be measured more accurately and faster than before mobile device popularization. To confirm this, we will compare the effect of measuring investor attention using search volume data before and after mobile device popularization. In addition, it is confirmed that the measured investor attention is that of retail traders, not institutional traders or professional traders, and the relationship between investor attention and short-term price pressure theory. Using SVI data provided by Google Trends, we will experiment with Russell 3000 stocks and IPO stocks and compare the results. In addition, the results of investigating the investor's interest using the search volume data from various angles through experiments such as the comparison of the results based on the inclusion of the noise ticker group, the comparison of the limitations of the existing investor attention measurement method, and the comparison of explanatory variables with existing IPO related studies. We would like to verify its practicality and significance.

# Contents

# Table of Contents

# Table of Figures

# Table of Definition

1. ASVI Variable

# Chapter 1. Introduction and Related Studies

　In this paper, it is confirmed whether the study of 'In Search of Attention' (2011) published in the Journal of Finance is also significant in the recent financial market. Existing research analyzed Google search volume data based on data up to 2008. However, if you look at the mobile device supply data released by Gartner as shown in [Figure 1] and Google's search volume data by year in [Figure 2], the explosive increase in mobile device usage and the resulting size of search data from 2009 You can see that it is much larger in the recent market. Against this background, the "In Search of Attention" study based on data from before 2008 cannot confirm whether the measure of investor interest using search data is significant even in the recent market where the amount of data has increased rapidly. In this study, the effect of the period before and after the popularization of smartphones is whether the measurement of the investor's interest level still using the search data is significant by using the search data that has rapidly increased and the recent market data due to the popularity of mobile devices, and whether there is price pressure depending on the level of investor interest. I want to compare.

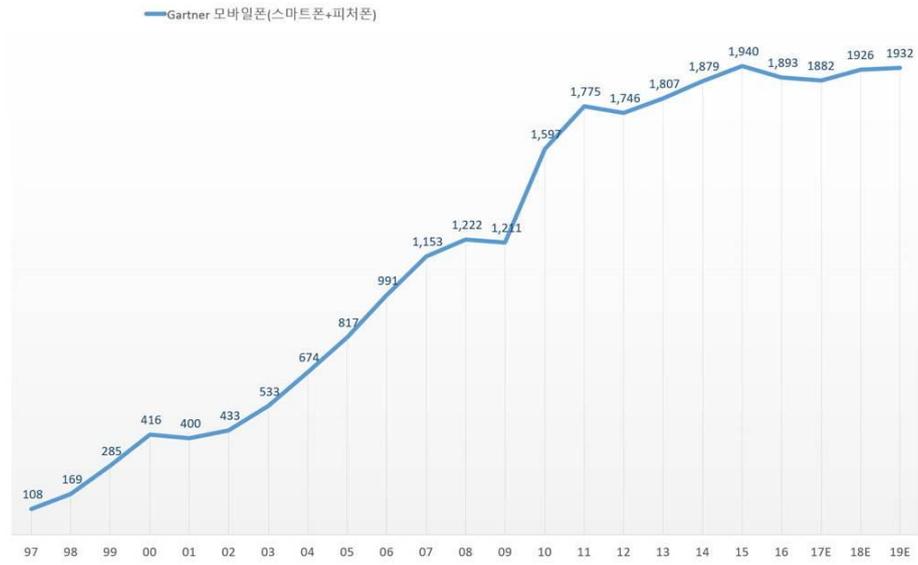

Figure 1: Gartner mobile phone supply

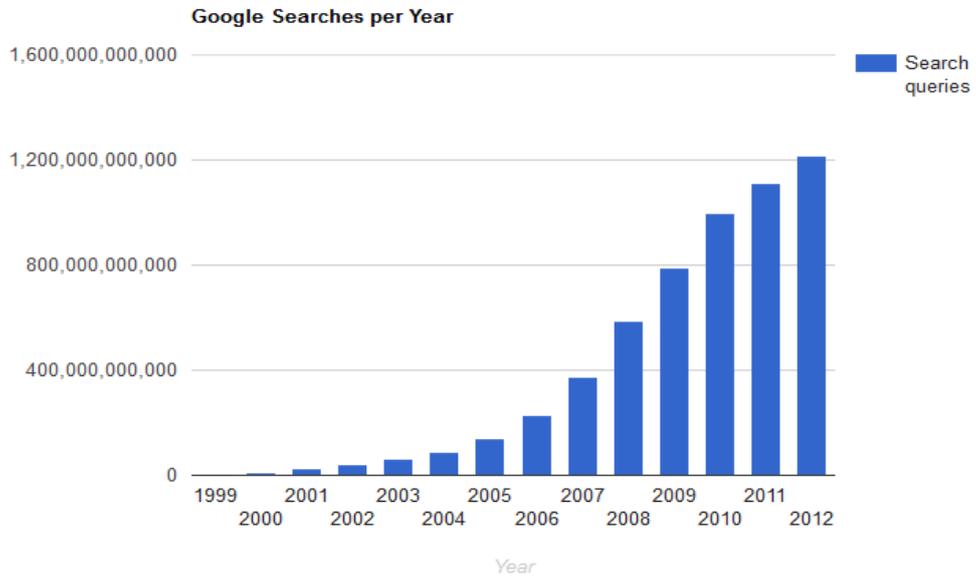

Figure 2: Annual Google search volume

# Chapter 2. Data and Variable Definition

## 2.1 Data Period and Range

 The Google Trends service provides data related to Google search information from January 2004. In this study, we use weekly Google search volume index data for each company. The stock group to be tested is the stock group and IPO stock group included in the Russell 3000 index. Russell 3000 is an index containing the top 3,000 companies in the United States, accounting for about 98% of stocks eligible for investments traded on the US stock market. The Russell 3000 is restructured annually to reflect market capitalization and reflects newly listed stocks that have grown rapidly that year. The Russell 3000's annual constituent list is taken from "Barcharts". In order to avoid survivor bias errors (Shumway (1997)) and reduce the effects of index inclusion and removal, all stocks included at least once in the sampling period are tested. Also, the IPO stock list is taken from Thomson Reuters and used. The period of all collected data is from 2004 to 2019. Starting in 2009, thanks to the popularization of smartphones and the spread of SNS, the size of search data has increased exponentially, and the speed at which information is delivered to investors has improved. Therefore, the data interval is divided into the period from 2004 to 2009, and from 2009 to 2019, and the difference in the experimental results for the two periods is compared.

## 2.2 Search keyword

 When using search data, it is important to select which search keyword to collect and use for experimentation. In general, investors use various search keywords such as company name, flagship product, and ticker symbol to obtain investment information. However, there are various problems when using a company name or flagship product as a search keyword.

First of all, there is a problem that when the company's name and service name are the same, such as 'Amazon', a search for the purpose of using the service rather than for investment purposes is included. In addition, there is a problem that various types of keywords can occur when the company name is used as a keyword. For example, if you are searching for the company name "American Airlines", you can search by several types of search keywords, such as "AMR Corp," "AMR," "American Airlines". In the case of searching for the company's flagship product, not only the same problems as in the above case, but also cases other than the purpose of searching for investment information such as product purchase purpose and product information search are included. Therefore, in this study, the company's ticker symbol is used as a search keyword to collect search data and use it in an experiment. In general, when the ticker symbol is used as a search keyword, it is highly likely that the search was used to obtain financial information and investment-related information of the company. In addition, since the ticker symbol is uniquely assigned to each company, there is an advantage in that companies can be clearly identified. However, when testing stocks before IPO, there is no ticker symbol because it is before listing on the exchange. Therefore, it is decided to use search data using the company name as a search keyword. In addition, there are cases in the ticker symbol that can be confused with other meanings other than the ticker symbol such as "DNA,", "GPS,", and "AA". We define these ticker symbols as noise ticker groups in advance, and check how the results differ when the noise group is removed and when the noise group is not removed in the robustness experiment. The noise group accounts for roughly 7-8% of Russell 3000 shares over the entire period.

## 2.3 Data collection

In order to collect search data of 3000 Russell stocks during the sample period, Google Trends uses web crawling. The web crawling program is implemented directly using Python.

After obtaining the SVI (Search Volume Index, hereinafter referred to as "SVI") using each company's ticker symbol as a search keyword, it is saved in CSV format and used. In Google Trend, if the search volume is smaller than the standard, the SVI value is assigned to 0. In this study, data with an SVI value of 0 are removed from the test subject. In order to compare the experimental results according to the search keyword, SVI (Name_SVI) using the company name as the search word and (PSVI) using the company's flagship product as the search word are also obtained. As for the company's flagship product, the top query in the related query provided in Google Trend is used as shown in [Figure 3]. Most of the top-level queries resulted in products or services provided by the company, but if the related query is not a product or service, it is used by manually matching the product or service among the ranking search terms.

**Figure 3: Query result related to 'APPL'**

The news data is from the Dow Jones Archives for news information relating to companies and IPOs, which were organized in Russell 3000 during the sample period. Since each article in the Dow Jones Archive is indexed as a set of company ticker symbols related to the article, it has the advantage of being easy to use by dividing each company. The IPO stocks subject to the experiment were those that were IPOs between 2004 and 2019, and the list of stocks IPOs during that period was taken from Thomson Reuter's data. Among them, closed funds, REITs, American Deposit Receipts (ADRs), Limited Partnerships (LPs), and stocks with a price of less than $5 are excluded. In addition, it will only test stocks that have traded on NYSE, Amex, and NASDAQ exchanges within 5 trading days after the IPO.

The Dash-5 report (SEC Rule 11Ac1-5) used to estimate the transaction history of individual investors in this study was taken from Thomson Reuter's data. Experiment based on the recorded content. Other data such as the company's price, transaction volume, company's financial information, and company-related analyst information used as basic data are taken from the Bloomberg terminal. [Table 1] is a table that predefined variables used in this study.

**Table 1: Variable definition**

|  | Name | Description |
|---|---|---|
| **related to Google Trends** | SVI | Frequency of Google searches by ticker symbol |
|  | ASVI | log(SVI) - log(Median(SVI for the previous 8 weeks) |
|  | NAME_SVI | Frequency of Google searches by company name |
|  | PSVI | Frequency of Google searches by product |
|  | APSVI | log(PSVI) - log(Median(PSVI for the previous 8 weeks) |
| **related to investor interest** | Abn Ret | Adjusted stock price return (Daniel 1997) |
|  | Abn Turnover | Normalized excess trading volume (Chordia 2007) |

| | |
|---|---|
| News | Dow Jones Archive Article Count |
| News Dummy | News Dummy Variable |

# Chapter 3. What is SVI?

## 3.1 Comparison with existing investor interest measures

SVI (Search Volume Index) is Google search volume index data provided by Google Trend. In this study, we would like to examine whether SVI can measure the interest of investors, especially individual investors. First, the correlation coefficient between the SVI and the value of the investor interest measurement replacement value used in the past is obtained as shown in [Table 2]. Each value used to calculate the correlation coefficient is based on weekly data and the period is the correlation coefficient for the entire period from 2004 to 2019.

**Table 2: Correlation coefficient with existing investor interest measure**

| | SVI | NAME_SVI | Abn Ret | Abn Turnover |
|---|---|---|---|---|
| **NAME_SVI** | 0.098 | | | |
| **Abn Ret** | 0.043 | 0.102 | | |
| **Abn Turnover** | 0.033 | 0.101 | 0.334 | |
| **News** | 0.05 | 0.145 | 0.201 | 0.176 |

Looking at the results of the correlation coefficient, in general, SVI shows positive correlation with the alternative value of other measures of investor interest, but the degree is less than that of other values. The correlation between Name_SVI and SVI appears to be around 9.8%. In the case of a company name, it can be used as a search keyword not only for investment purposes, but also for various reasons such as searching for a company location or purchasing a product, but most of the company's ticker symbols are likely to be used as search keywords only for investment purposes. Because of this size. Absolute Abn Ret and Abn Turnover are alternative values for investor interest measurements based on transaction data that were commonly used before the SVI. The two have a correlation of more than 30% with each other, but have a relatively low correlation, such as 3.3 to 4.3%, with SVI, an alternative value for measuring new interest. This is because the absolute value of the excess return and the excess transaction volume are values extracted from the transaction data, and the transaction has a lot of influence not only on investor interest, but also on other economic factors.

News data is also a measure of investor interest, which has been widely used. Looking at the correlation coefficient, the amount of news data also has a positive correlation with SVI, but the degree is low. However, there are some problems with news data. First of all, news data doesn't cover most of the stocks in Russell 3000. In addition, there is a problem that it cannot be guaranteed that it is the interest of the investor until the investor directly reads the news related to the company. In addition, there are news that the company-related news affects investors, but because the name or symbol of the company is mixed for marketing or comparison with other companies, it is difficult to say that the article is a pure measure of investor interest. On the other hand, SVI has very small search data, so it can measure the interest of most equity investors in Russell 3000 right away, except for a few stocks with an SVI value of zero. Of course, due to the advent of the news, search increases and SVI may increase temporarily, but such news is news that directly affects investors, and SVI can be used as an alternative measure of investor interest.

## 3.2 VAR weekly delay relationship analysis

Next, we analyze the weekly delay relationship between investor interest measurement replacement values using VAR (vector autoregression). The data of each value is the data interval after 2009, and weekly data of at least 2 years is used. For the coefficients shown in [Table 3], the VAR coefficient is calculated for each stock and the average value is calculated at each time point. The p-value in parentheses in [Table 3] is the hypothesis that all of the VAR coefficients are zero, that is, the hypothesis that the weekly delayed investor interest measurement replacement values are not related to the replacement values at the present time. (2006)). In the block bootstrap method, it is important to determine the size of the block. The block length used in this study uses 23 weeks to maintain a certain level of autocorrelation.

The experimental results show that SVI appears one week ahead of the existing alternative measures of investor interest (absolute value of excess return, excess trading volume, and news volume). When experimenting with alternative values of interest in other investors as dependent variables of VAR, the coefficients of the weekly delay log (SVI) are all positive values. In other words, the weekly delayed SVI well explains the replacement values of existing investor interest, which shows that SVI can measure investor interest at an earlier time. The reason for this result is that the absolute value of the excess return that was used as an alternative value for investor interest and the excess trading volume are values obtained through transaction data, so they are measured after the transaction occurs. The above result is shown because the transaction is preceded by investor information search and an increase in investor interest prior to trading. In addition, since news data and investor interest increase, reporters will analyze and write articles with interest, so investors' interest increases before the news appears, and SVI measures it well in advance.

If you place Log (SVI) as the dependent variable, Log (turnover) and Log (1+News) have significantly negative coefficient values. This is because after news or high excess trading volume, SVI returns to average and falls again. On the other hand, the absolute value of the excess return has significantly positive coefficient values, because the investor's interest continues even after a high excess return. As such, SVI can measure investor interest earlier than existing alternative values.

**Table 3: VAR model of investor interest measurement proxies**

| | Lagged 1 Week | | | | |
| --- | --- | --- | --- | --- | --- |
| | SVI | Abn Turnover | Abn Ret | log(1+News) | R^2 |
| **SVI** | 0.5764*** | -0.0012*** | 0.0512*** | -0.0016*** | 59.45% |
| | (0.01) | (0.01) | (0.01) | (0.01) | (0.01) |
| **Abn Turnover** | 0.0512** | 0.4976*** | 0.5092*** | -0.0401*** | 32.89% |
| | (0.05) | (0.01) | (0.01) | (0.01) | (0.01) |
| **Abn Ret** | 0.0037*** | 0.0009*** | 0.0405*** | -0.0008*** | 3.67% |
| | (0.01) | (0.01) | (0.01) | (0.01) | (0.05) |
| **log(1+News)** | 0.0712** | 0.0192*** | 0.2283** | 0.0201*** | 2.91% |
| | (0.02) | (0.01) | (0.05) | (0.01) | (0.01) |

## 3.3 ASVI Variable

In this study, in order to accurately measure the level of interest of a special investor adjusted for the level of interest of investors in the normal range rather than the level of interest of investors in the normal range, the SVI value is not used immediately, but a new variable called ASVI is created and used. ASVI refers to the value obtained by subtracting the value taken from Log (SVI) at time t from the median value of SVI for the previous 8 weeks as shown in [Equation 1]. The ASVI adjusted its degree because the median SVI for eight weeks could be viewed as a normal range of investor interest. And ASVI has the effect of removing trends and seasonality.

**Definition 1: ASVI Variable**

$$ASVI_t = \log(SVI_t) - \log[Med(SVI_{t-1}, \ldots, SVI_{t-8})]$$

# Chapter 4. Analysis of the relationship between SVI and retail investor interest

## 4.1 Retail Investor Transaction

If you think intuitively, investors who search for investment information on the Internet (Google) are likely to be individual investors, not institutional investors or professional investors. Institutional investors or professional investors can use specialized terminals such as Bloomberg or Reuters, analyst seminars, etc. to obtain or search investment information, but individual investors cannot use these specialized terminals, so they can use personal computers or mobile devices. Search the Internet for investment information. In this study, we will find out whether SVI is actually a substitute value for investor interest, which can measure the interest of investors, especially individual investors.

Existing studies (Han (2013)) generally use ISSM or TAQ databases to extract individual investors' transaction details and use them as data in order to obtain transaction details of individual investors. However, the strategy for institutional and professional investors to divide orders began to become active (Caglio and Myhew (2008)), and after 2001, order transaction quantities began to be expressed in decimals. Research results are known that it has become difficult to properly extract the transaction details of investors. (Hvidkjaer (2008)). Considering that the sample period used in this study is data from 2004 to 2019, it is not appropriate to extract transaction details of individual investors using ISSM or TAQ databases. In this study, Dash -5 Use the monthly report.

Since 2001, the SEC has required all exchanges in the US to submit monthly reports related to "Covered orders" in accordance with Rule 11Ac1-5 and Regulation 605. Covered orders are transaction details excluding special orders that require special manipulation and excluding large orders over 10,000 shares. For this reason, it can be seen that it is mainly composed of the transaction details of individual investors who do not require special manipulation and have a small transaction size. In fact, a study by Boehmer, Jennings, and Wei (2007) revealed that most of the transactions included in the Dash-5 report occurred as individual investors. In the Dash-5 report, the order quantity is divided into four categories: (1) 100-499 (2)500-1,999 (3)2,000-4,999 (4)5,000-9,999. This order volume can be used to calculate the individual investor's monthly trading volume and change in turnover rate. And experiment the relationship between the calculated change values and the monthly SVI change. Monthly SVI is the sum of the weekly SVIs included in the month. Also, compare other values of interest measurement alternatives. The sampling period is compared by dividing the years 2004 to 2008 and 2009 to 2019. And since small orders (100-1,999 shares) can better represent the transaction details of individual investors, the experiment results are compared by dividing them by the number of orders.

**Table 4: Relationship between interest measurement value and retail investor transaction**

| Periods | 2004-2008 | | | |
|---|---|---|---|---|
| Order Size | Order Size: 100- 1,999 | | Order Size: 100- 9,999 | |
| | ΔOrder | ΔTurnover | ΔOrder | ΔTurnover |
| ΔSVI(t-1,t) | 0.0925*** | 0.0920*** | 0.103*** | 0.133*** |
| | (0.01) | (0.00924) | (0.0108) | (0.0119) |
| Ret(t) | 0.111*** | 0.123*** | 0.0988*** | 0.00863 |
| | (0.0258) | (0.0233) | (0.0268) | (0.0232) |
| Abn Ret(t) | 0.901*** | 1.011*** | 1.047*** | 1.592*** |
| | (0.041) | (0.0443) | (0.05) | (0.0512) |
| News Dummy(t) | 0.0781*** | 0.0932*** | 0.0925*** | 0.119*** |
| | (0.003) | (0.00283) | (0.00302) | (0.00321) |
| Constant | 0.129*** | 0.139*** | 0.155*** | 0.1802*** |
| | (0.0145) | (0.0145) | (0.0145) | (0.0123) |
| R^2 | 0.24 | 0.275 | 0.269 | 0.298 |
| Periods | 2009-2019 | | | |
| Order Size | Order Size: 100- 1,999 | | Order Size: 100- 9,999 | |
| | ΔOrder | ΔTurnover | ΔOrder | ΔTurnover |
| ΔSVI(t-1,t) | 0.1594*** | 0.1493*** | 0.1943*** | 0.2113*** |
| | (0.012) | (0.011) | (0.0132) | (0.018) |
| Ret(t) | 0.129*** | 0.122*** | 0.1293*** | 0.00983* |
| | (0.019) | (0.0212) | (0.0194) | (0.020) |
| Abn Ret(t) | 1.102*** | 1.293*** | 0.927*** | 1.4932*** |
| | (0.049) | (0.022) | (0.04) | (0.0483) |
| News Dummy(t) | 0.0429*** | 0.0762*** | 0.0458*** | 0.0982*** |
| | (0.002) | (0.00199) | (0.0031) | (0.00276) |
| Constant | 0.112*** | 0.129*** | 0.132*** | 0.1792*** |
| | (0.0231) | (0.0201) | (0.0198) | (0.0213) |
| R^2 | 0.302 | 0.327 | 0.325 | 0.339 |

## 4.2 Relationship between retail investor transactions and SVI

First, let's check the experimental results of the data section from 2004 to 2008 in [Table 4]. Looking at the relationship between the change in the number of orders and the change in SVI, when the SVI changes by 1%, the number of orders of individual investors changes by 0.0925%. This significant positive relationship is similarly significant when other interests are adjusted for substitution values. When looking at the relationship between the absolute value of the stock price return and excess return and the change in the number of orders, it shows a significant positive relationship with the change in the number of orders, like the change in SVI. However, this result is hard to be seen as a relationship derived purely from investor interest because the values used as substitute values for investor interest are themselves extracted through calculation formulas from transaction data. In addition, an experiment with the change in transaction turnover as a dependent variable shows a similar appearance to the result of setting the change in the number of orders as a dependent variable. Dividing the size of the number of orders into 100-1,999 weeks and 100-9,999 weeks, and comparing groups with a small number of orders separately, results in similar results without much difference between the two. Therefore, it can be seen that the Dash-5 report limits individual investor transactions to less than 2,000 shares, so there is no need to compare them.

Next, let's check the results of experimenting the relationship between interest measurement substitution values and individual investor transactions in the data period from 2009 to 2019 in [Table 4]. First, looking at the relationship between the change in the number of orders and the change in SVI, when the SVI changes by 1%, the number of orders of individual investors changes by 0.1594%. This significant positive relationship, like the data period from 2004 to 2008, not only shows a significant positive relationship, but also shows a greater degree of relationship in the recent period than in the past period. In other words, in the period after 2009, SVI is better explaining orders from individual investors. In an experiment with the amount of change in the transaction turnover as a dependent variable, as in the data period from 2004 to 2008, not only does it show a significant positive relationship, but also shows a large degree of relationship. This is because, with the proliferation of mobile devices after 2009, as more investors use the search service, the SVI value of the recent period can better explain the transaction details of individual investors than the past period.

Looking at the replacement values of the SVI change as well as the existing investor interest, the overall value does not change significantly compared to the period before 2009. Since stock price returns and excess returns are values extracted from transaction data, the increase in search data after 2009 was not affected by the change of the data period to the latest period. In the case of news data, the relationship slightly decreases in the period after 2009. This can be seen as a decrease in the degree of measuring investor interest as article data that is not related to investment or company fundamentals increases, such as for the company's marketing purposes.

# Chapter 5. Analysis of the relationship between SVI and price pressure

## 5.1 Price pressure

Barber and Odean (2008) study confirmed that individual investors do not have any special restrictions when buying stocks, but rather have various options, but there are restrictions such as having to already own stocks or being eligible for short selling when selling. I did. Therefore, existing studies have argued that when individual investors' interest in a particular stock increases, the weakly constrained buy tax becomes stronger than the highly constrained sell tax, and a bubble in the price temporarily increases the short-term stock price return.

In this study, after measuring the level of interest of individual investors using ASVI, the price pressure phenomenon according to the level of investor interest identified in previous studies is verified using the latest market data.

First, we look at the relationship between ASVI and price pressure for stocks in Russell 3000 and then analyze the relationship to price pressure for IPO stocks. In the case of IPO stocks, there is no transaction data prior to listing, so it is not possible to use alternative values for measuring existing interest levels such as excess returns or volume calculated from transaction data. However, even before listing, investors can express their interest in IPO stocks through search, so measuring the interest level using SVI is the only data that can be used to analyze how investors' interest affects IPO stocks.

## 5.2 Russell 3000

In this study, Fama-MacBeth (1973) cross-sectional analysis methodology is used. First, for each week, the DGTW excess return (Daniel 1997) of the next week (weeks 1, 2, 3, 4, 5-52) is calculated as a basis point and then used as a dependent variable. The coefficient values in [Table 5] are the averaged values of individual stocks. The standard deviation is calculated using the Newey-West (1987) formula.

Table 5: ASVI and Russell 3000 Fama-MacBeth Cross Section Analysis

| Periods | 2004-2008 | | | | |
|---|---|---|---|---|---|
| | Week1 | Week2 | Week3 | Week4 | Week 5-52 |
| ASVI | 18.741 *** | 14,904** | 3.85 | -1.608 | -28.912 |
| | (7.01) | (7.56) | (6.284) | (6.903) | (17.162) |
| Log Mkt Cap x ASVI | -21.182*** | -15.467** | -4.71 | 4.29 | 16.834 |
| | (6.508) | (6.768) | (6.516) | (6.398) | (88.624) |
| Log Mkt Cap | 2.653 | 3.858 | 3.144 | 3.557 | -39.229 |
| | (3.023) | (3.160) | (3.062) | (3.186) | (67.405) |
| Percent Dash-5 Volumn x ASVI | 3,552** | 1.904 | 1.687 | -2.744 | 16.258 |
| | (1.639) | (1.522) | (1.612) | (1.717) | (23.822) |
| Percent Dash-5 Volumn | 1.607 | 1.351 | 1.486 | 0.364 | 119.901*** |
| | (1.644) | (1.652) | (1.659) | (1.711) | (31.765) |
| APSVI | -2.532*** | -1.387 | -0.701 | -0.704 | 2.286 |
| | (0.930) | (0.98) | (0.808) | (0.638) | (9.909) |
| Abn Ret | 1.314 | -2.389 | -1.128 | -0.463 | -1.510 |
| | (1.879) | (1.979) | (1.563) | (1.405) | (28.505) |
| News Dummy | 3.610* | 1.387 | -3.825 | -0.058 | 32.466 |
| | (2.025) | (2.424) | (2.483) | (1.910) | (28.440) |
| Abn Turnover | 2.393** | 2.309** | 2.022 | 0.316 | 10.531 |
| | (1.204) | (1.144) | (1.404) | (1.098) | (10.109) |
| R^2 | 0.0142 | 0.0119 | 0.0112 | 0.0111 | 0.017 |
| Periods | 2009-2019 | | | | |
| | Week1 | Week2 | Week3 | Week4 | Week 5-52 |
| ASVI | 24.392 *** | 17.923** | 2.937 | -2.837 | -31.283 |
| | (5.64) | (6.94) | (5.947) | (6.943) | (18.273) |
| Log Mkt Cap x ASVI | -22;983*** | -14.923** | -3.92 | 5.392 | 17.923 |
| | (7.028) | (6.331) | (5.927) | (6.928) | (90.271) |
| Log Mkt Cap | 3.023 | 4.0293 | 3.872 | 4.092 | -41.927 |
| | (2.997) | (3.213) | (2.981) | (3.989) | (66.908) |
| Percent Dash-5 Volumn x ASVI | 3.821** | 2.018 | 1.729 | -2.429 | 17.928 |
| | (1.829) | (1.239) | (1.521) | (1.829) | (22.382) |
| Percent Dash-5 Volumn | 1.782 | 1.472 | 1.232 | 0.413 | 132.392*** |
| | (1.082) | (1.5372) | (1.718) | (1.927) | (34.273) |
| APSVI | 5.292*** | 3.293 | 0.879 | 2.392 | 4.593 |
| | (0.873) | (0.872) | (0.901) | (0.823) | (10.372) |
| Abn Ret | 1.928 | -2.193 | -1.827 | -0.283 | -1.927 |
| | (2.091) | (1.826) | (1.891) | (1.832) | (30.283) |
| News Dummy | 6.593** | 2.327 | -4.293 | 0.082 | 36.293 |
| | (3.282) | (2.012) | (3.293) | (2.018) | (30.827) |
| Abn Turnover | 2.219** | 2.1392** | 1.927 | 0.219 | 9.729 |
| | (1.402) | (1.023) | (1.293) | (0.973) | (11.283) |
| R^2 | 0.0203 | 0.0192 | 0.0189 | 0.0186 | 0.0201 |

First of all, looking at the results of the data period from 2004 to 2008 in [Table 5], ASVI has a statistically significant and positive relationship with the excess return in the next week 1 or 2 period. Since each independent variable uses a standardized value, the regression coefficient can be interpreted as the influence of the change of 1 standard deviation. For example, an increase of 1 standard deviation of ASVI can be interpreted as causing a price change of +18.7 bps on average within the Russell 3000 stock. And if this price increase is due to the buying trend of individual investors rather than other rising factors, the trend will be more pronounced in stocks with small market caps, where individual investors are actively buying. In fact, if you check the coefficient of the product of market capitalization and ASVI, you can see that it has a significantly large negative coefficient value. In other words, it can be seen that most of the price increases in the first and second weeks are seen in stocks with small market caps. In addition, if you look at the Dash-5 ratio variable, which represents the transaction details of individual investors, divided by the total transaction volume by the Dash-5 transaction volume, the product of the Dash-5 ratio and ASVI has a significant positive coefficient in week 1. In other words, it can be seen that the price increase in the early period was driven by the transaction of individual investors.

Looking at the results for weeks 2-3, we can see that ASVI is still explaining the price increase. In week 2, an increase of 1 standard deviation of SVI induces a price change by +14.9 bps, and in week 3, a change by +3.85 bps. However, after week 4, the price pressure caused by investor interest decreases, and the stock price begins to rebound. In week 4, an increase of 1 standard deviation of SVI caused a price change of -1.6 pbs, and from week 5 to week 52, an increase of 1 standard deviation of SVI caused a price change of -28.9 bps. Cause. Price pressure caused by investor interest shows that for the first one to two weeks, the price increase is triggered by price pressure due to investor interest, but most of the price increases revert within one year. Also, if you check the results of APSVI, which uses the company's main product as a search keyword, it has significant negative coefficient values in the first week or two. It can be expected that the initial price increase was not due to the launch of new products, innovation of flagship products, or the recovery of the company's own fundamentals, but mostly due to price pressure caused by investor interest. Looking at the results of alternative values of the existing interest in other investors, it is not possible to properly explain the temporary price rise pressure caused by investor interest in the study of Barber and Odean (2008).

Next, let's look at the results of the data period from 2009 to 2019 in [Table 5]. Overall, the results of the pre-2009 sampling period and the relationship are similar, but the degree of relationship between each is larger. Looking at the relationship between the excess return and SVI for the next week 1, it has a significantly positive relationship. It can be seen that an increase of 1 standard deviation of ASVI significantly causes a change in average price of +24.392 bps in Russell 3000 stocks, more than that caused by a price change of 18.7 bps in the pre-2009 data period. Appear large. If we compare the long-term regression trend of stock prices as the price pressure caused by investor interest disappears, we can see that stock prices rebound more significantly earlier than in the period before 2009. It seems that this is because information about the company is delivered to investors more quickly due to the advent of mobile devices and the expansion of search services after 2009, and investors search for more investment information through search. In other words, it can be said that the speed of acquiring information and making investment decisions faster than in the past. In addition, with the advent of the HTS (Home Trading System) and the ease of use of brokerage services using mobile, the shorter investment cycle for investors can be seen as the reason for the faster return.

Also, if you look at the news data, you can see that the predictive power is even greater than in the past. Although there was noise news, it is interpreted that the spread of the Internet and the advent of mobile devices accelerated the spread of news to investors. Looking at the coefficients of APSVI, it can be seen that the explanatory power of over-period returns is improved in the data period after 2009 compared to the data period before 2009. This can be interpreted in two aspects. First of all, as the name of the company's flagship products and services, such as Facebook and Twitter, has become more common in the recent market, the rate of search for investment information has increased even when a search keyword is used as a flagship product. to be. And with the popularization of mobile devices, the speed of information delivery has accelerated, allowing investors to quickly recognize the company name. As a result, fundamental parts such as the company's flagship product spread to investors more quickly, and the explanatory power of the keyword search for the flagship product improved. As such, in the period after 2009, even though the APSVI variable has a large positive significance, the significance of ASVI does not disappear or decrease in size, but rather has a greater explanatory power. I can confirm. Looking at the coefficient of the Dash-5 ratio, it has a significantly positive coefficient as in the pre-2009 period. This is because individual investors' transactions led the initial price increase after 2009 as well.

### 5.2.1 Noise ticker group

Next, let's check the experimental results for each data period after removing noise tickers such as "GPS", "DNA", "BABY" and "A" of the entire sampling period. Noise tickers account for about 7.4% of the total tickers.

**Table 6: ASVI and Russell 3000 Fama-MacBeth Cross Section Analysis-Noise Ticker Removal**

| Periods | 2004-2008 | | | | |
|---|---|---|---|---|---|
| | Week1 | Week2 | Week3 | Week4 | Week 5-52 |
| ASVI | 18.294*** | 15.293** | -0.684 | -6.1293 | -28.283 |
| | (8.213) | (8.271) | (7.212) | (7.423) | (19.382) |
| Log Mkt Cap x ASVI | -20.829*** | -16.182** | -0.321 | 8.212 | 12.381 |
| | (7.972) | (7.281) | (7.573) | (7.012) | (78.928) |
| Log Mkt Cap | 3.421 | 4.602 | 3.682 | 3.919 | 13.281 |
| | (2.892) | (3.001) | (2.829) | (3.281) | (50.231) |
| Percent Dash-5 Volumn x ASVI | 3.011** | 1.201 | 0.917 | -3.452* | 16.829 |
| | (1.573) | (1.702) | (1.921) | (1.621) | (22.281) |
| Percent Dash-5 Volumn | 1.110 | 0.699 | 0.829 | -0.401 | 82.392*** |
| | (1.631) | (1.702) | (1.722) | (2.019) | (29.728) |
| APSVI | -1.906*** | -1.082 | 0.256 | 0.352 | 1.201 |
| | (0.901) | (0.998) | (0.813) | (0.601) | (8.432) |
| Abn Ret | 2.012 | -3.018 | -1.173 | -1.312 | -12.201 |
| | (2.121) | (2.129) | (1.701) | (1.892) | (15.012) |
| News Dummy | 2.091* | 0.501 | -3.321 | -0.812 | 15.782 |
| | (2.231) | (2.912) | (2.501) | (2.212) | (23.876) |
| Abn Turnover | 1.978 | 2.581** | 3.012** | 0.398 | 20.812 |
| | (1.312) | (1.291) | (1.453) | (1.321) | (10.874) |
| R^2 | 0.0155 | 0.0118 | 0.0119 | 0.0116 | 0.018 |
| **Periods** | 2009-2019 | | | | |
| | **Week1** | **Week2** | **Week3** | **Week4** | **Week 5-52** |
| ASVI | 19.287*** | 17.032** | -0.781 | -6.392 | -29.281 |
| | (8.301) | (8.271) | (7.342) | (7.411) | (20.291) |
| Log Mkt Cap x ASVI | -21.232*** | -15.393** | -1.291 | 7.593 | 13.293 |
| | (8.022) | (7.123) | (7.231) | (7.942) | (92.183) |
| Log Mkt Cap | 3.523 | 4.232 | 3.912 | 3.534 | 14.232 |
| | (2.018) | (3.293) | (2.431) | (3.102) | (50.231) |
| Percent Dash-5 Volumn x ASVI | 3.231** | 1.403 | 1.028 | -3.998* | 17.920 |
| | (1.692) | (1.802) | (2.082) | (1.826) | (24.938) |
| Percent Dash-5 Volumn | 1.203 | 0.826 | 0.972 | -0.382 | 84.382*** |
| | (1.729) | (1.803) | (1.982) | (1.927) | (30.729) |
| APSVI | 4.892*** | -0.928 | 0.274 | 0.421 | 1.392 |
| | (0.972) | (0.921) | (0.892) | (0.679) | (8.728) |
| Abn Ret | 2.195 | -2.938 | -1.092 | -1.293 | -11.928 |
| | (2.392) | (2.098) | (1.866) | (1.782) | (14.928) |
| News Dummy | 3.083* | 0.683 | -3.192 | -0.728 | 16.922 |
| | (2.242) | (2.899) | (2.711) | (2.432) | (24.928) |
| Abn Turnover | 2.073 | 2.692** | 3.129** | 0.402 | 21.392 |
| | (1.499) | (1.302) | (1.582) | (1.492) | (11.028) |
| R^2 | 0.0160 | 0.0121 | 0.0123 | 0.0119 | 0.020 |

Looking at the results after excluding the noise ticker group in [Table 6], both before and after 2009, ASVI still explained the price increase in the first two weeks well, especially in stocks with a small market cap and a lot of private investors Appear large. In addition, most of the price increases caused by price pressure return within one year. Other investor interest substitution values other than SVI do not show this result. The noise ticker group accounts for about 7.4% of the total data on average, but it does not significantly affect the results when included or removed.

## 5.3 IPO stock

## 5.3.1 IPO stock's Return

In fact, the best group to see the impact of individual investor interest on stock prices is the stock group that has recently become an IPO. According to previous studies, on average, IPO stocks are known to exhibit high stock price returns on the first day of listing (Loughran 2002). However, stocks that showed a high stock price return on the first day tend to show a low return in the long run due to a regression in stock prices. (Loughran 1995). Considering these characteristics, the hypothesis of pressure to raise prices caused by investor interest in the previous Barber and Odean (2008) study can naturally be applied to the IPO stock group recently. IPO stocks have many opportunities to receive a lot of attention from investors even before they are listed on the exchange due to issues such as an announcement of a listing plan and an announcement of a target price from an unlisted company. And investors want to search companies for information whenever these issues arise. Therefore, it is highly likely that stocks that have received high investor interest from the time of listing are subject to high buying pressure from investors on the first day of trading on the actual exchange. As time passes after the listing, interest and price pressure in the early stage fall, leading to a long-term retracement of the stock price, leading to a low stock price return.

Existing studies have explained the phenomena of IPO stocks showing high stock price returns on the first day of listing and low price returns for a long period based on investor sentiment data (Ritter and Welch (2002), Ljungqvist, Nanda, and Singh (2006), and Cook , Kieschnick, and Van Ness (2006)). According to existing studies, if investors' sentiment about IPO stocks intensifies, the stock price is likely to be higher than the normal price due to the pressure on the first day of listing on the exchange. As they return, price pressures gradually weaken and return to normal prices in the long run. In general, the degree of sentiment of investors before IPO can be extracted from the IPO listing price determined before listing. When investor expectations are high and sentiment begins to intensify, more investors will want to participate in IPOs, and IPO listing prices will rise as demand increases. In fact, according to Cornelli, Goldreich, and Ljungqvist (2006) study, IPO listing price has a significant positive correlation with the stock price return of the IPO stock on the first day, and a significant negative correlation with the stock price return of the IPO stock after 52 weeks. I can see it. In other words, the stock price return on the first day is under price pressure due to investor sentiment, and over time, as the sentiment returns to the normal level, price pressure weakens and returns to the normal price.

However, in order for investors to develop big and small feelings about a particular stock, investors' attention is first required. That's because you need to be interested in creating feelings for the stock. The intensification of investors' sentiment also means that investors' interest has already increased before. Therefore, the interest of the investor and the sentiment of the investor are bound to show a positive correlation. In other words, using ASVI, it is possible to measure not only the interest of the investor but also the emotion of the investor.

## 5.3.2 SVI in IPO stock

First you need to find the SVI of IPO stocks. However, stocks prior to the IPO do not have a ticker symbol because they are not listed on the exchange. Therefore, in the case of stocks before IPO, a new SVI using the company name as a search keyword is obtained and used in the experiment. The sampling period is from 2009 to 2019, and the IPO was conducted for the entire period, and the experiment is conducted on 971 significant stocks whose SVI value provided by Google Trend is not 0.

First, let's look at how the SVI value changes before and after the IPO. [Figure 4] is a graph showing the average and median values of SVI by week and ASVI by week in log scale. Looking at the graph, both the average and median values have increased overall SVI from two weeks before the IPO, and have the highest SVI values in the IPO week. After the IPO, the SVI value gradually declines. In fact, in the 2-3 weeks prior to the IPO and during the week of the IPO, the interest of investors will increase because a lot of IPO stock promotions and related articles are published to investors. Like SVI, if you look at ASVI, which is a value adjusted for interest in the normal range, on average during the IPO week, the ASVI is about 28% higher than the previous period.

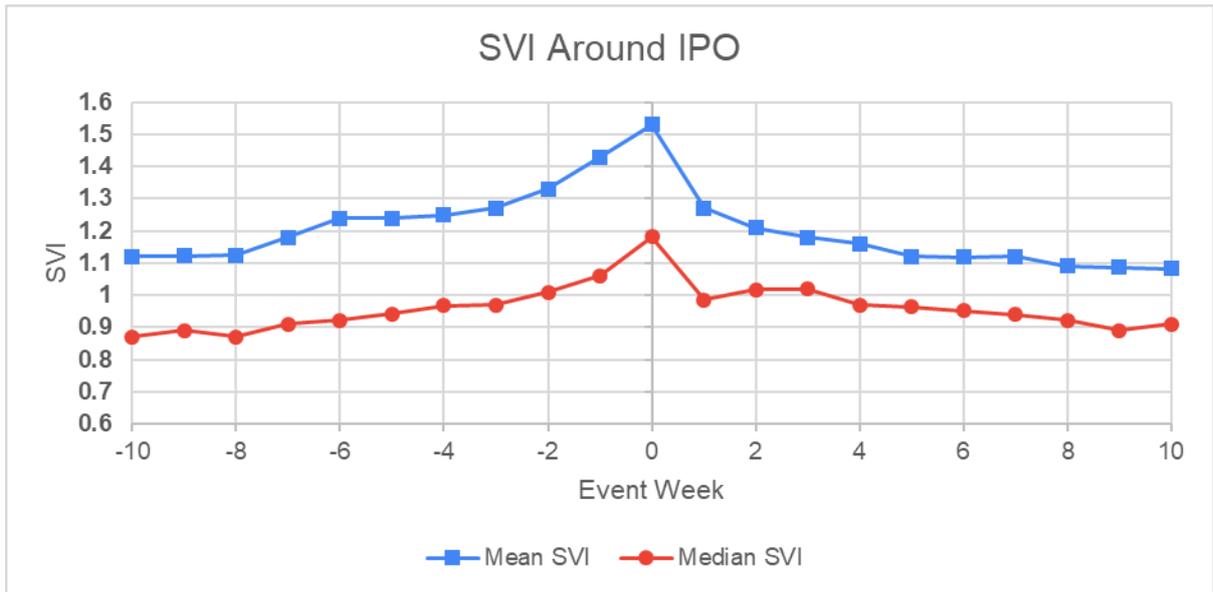

**Figure 4: SVI during IPO**

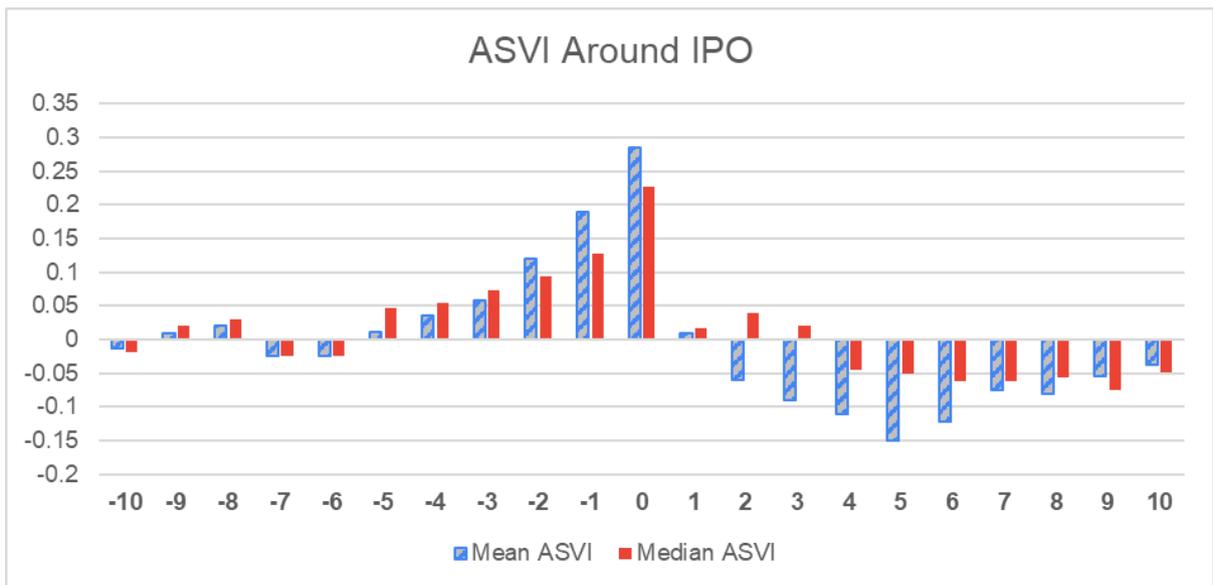

**Figure 5: ASVI during IPO**

Next, let's look at the relationship between the rise in investor interest before the IPO and the stock price return on the first day of the IPO. The results of [Table 7] and [Figure 6] support the assertion that the stock price rises high on the first day, triggered by investor interest in the existing studies. The average return on the first day of stocks with low ASVI during the first week before the IPO was about 11.73%, while the average on the first day of stocks with high ASVI during the pre-IPO period was about 17.64%. The difference in average return was 5.91%, which was found to be a statistically significant difference in the 1% interval as a result of the t-test. Also, the median comparison of stocks that showed high ASVI over the previous week showed higher returns on the first day. However, when looking at the returns over the 5-52 weeks, on the contrary, the stocks that showed high ASVI during the week before the IPO show low returns in both the average and median values. This seems to be due to the price pressure from increased investor interest rather than the fundamentals of the company, rather than the rise in prices at the beginning of IPO listing. Over time, the level of interest decreases, and the stock price returns accordingly.

**Table 7: First day return by ASVI before IPO**

| Periods | ASVI before IPO | Average return | Median return |
|---|---|---|---|
| IPO first day | Low | 11.73% | 5.95% |
| | High | 17.64% | 11.38% |
| 5-52 weeks after IPO | Low | 4.53% | 1.45% |
| | High | -2.10% | -22.83% |

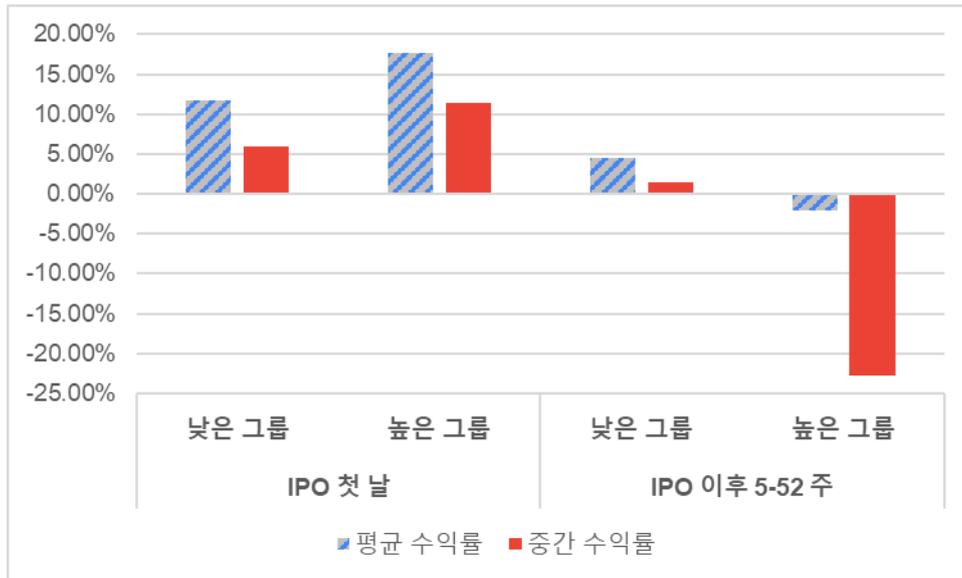

**Figure 6: First day return by ASVI before IPO**

## 5.3.3 Relationship between IPO stock's return and prime variables

Next, we experiment with regression analysis using the average stock price return on the first day of individual IPO stocks as a dependent variable. As an independent variable, two independent variables, which have high predictive power of the first day of IPO stocks, are additionally tested, as well as ASVI, which is a substitute value for investor interest to be verified in this paper.

The first variable is the media variable (Liu (2009)). This variable is a logarithmic scale of the total number of news articles that appear when a company name is searched from the date of IPO document submission to the day before the actual listing on the exchange (Cook, Kieschnick, and Van Ness (2006)). The media variable is one of the substitute values not only for predicting the stock price return of IPO stocks on the first day, but also for investor interest that was used previously.

The second variable is the price correction variable. This variable is the ratio obtained by dividing the final public offering price listed by the median of the range of the desired offering price submitted as a document. According to a previous study (Hanley (1993)), the larger the price correction rate, the higher the return on the first day. The high price correction ratio means that the final offering price is higher than the desired offering price. This also means that there is high demand from investors who want to participate in IPOs. In addition, in order to confirm the results, the results are experimentally compared by controlling the characteristic variables related to the IPO company (asset size), industry (industrial rate of return), and IPO (procurement size). And the sampling period is divided into the period before and after 2009, and the results are compared in the latest market where search data has increased rapidly.

**Table 8: Relationship to returns on the first day of the IPO**

| Periods | 2004-2008 | | | | | | |
|---|---|---|---|---|---|---|---|
| ASVI | 0.276** | | | 0.204** | | | 0.188** |
| | (0.112) | | | (0.0786) | | | (0.0712) |
| Media | | 0.0301* | | | 0.0255** | | 0.0244 |
| | | (0.0189) | | | (0.0113) | | (0.0134) |
| Price Revision | | | 0.460*** | | | 0.365*** | 0.351*** |
| | | | (0.0812) | | | (0.101) | (0.101) |
| Offering Size | | | | 0.0804*** | 0.0722*** | 0.0343 | 0.0167 |
| | | | | (0.013) | (0.0127) | (0.0219) | (0.0176) |
| Asset Size | | | | -0.0452*** | -0.0446*** | -0.0237*** | -0.0199** |
| | | | | (0.00987) | (0.00975) | (0.00801) | (0.00591) |
| Industry Return | | | | 0.201** | 0.258*** | 0.127 | 0.184** |
| | | | | (0.0901) | (0.0743) | (0.102) | (0.0854) |
| Constant | 0.113*** | 0.0540 | 0.143*** | -0.747*** | -0.713*** | -0.302 | -0.179 |
| | (0.0146) | (0.0408) | (0.0126) | (0.184) | (0.180) | (0.274) | (0.222) |
| $R^2$ | 0.052 | 0.035 | 0.235 | 0.217 | 0.215 | 0.289 | 0.341 |

| Periods | 2009-2019 | | | | | | |
|---|---|---|---|---|---|---|---|
| ASVI | 0.321** | | | 0.279** | | | 0.219** |
|  | (0.188) | | | (0.1029) | | | (0.0815) |
| Media | | 0.0201* | | | 0.0198** | | 0.0208 |
|  | | (0.0117) | | | (0.0097) | | (0.0127) |
| Price Revision | | | 0.448*** | | | 0.343*** | 0.339*** |
|  | | | (0.0791) | | | (0.098) | (0.097) |
| Log(Offering Size) | | | | 0.0698*** | 0.0658*** | 0.0298 | 0.0155 |
|  | | | | (0.009) | (0.0119) | (0.0201) | (0.0143) |
| Log(Asset Size) | | | | -0.0318*** | -0.0401*** | -0.0201*** | -0.0178** |
|  | | | | (0.00879) | (0.00952) | (0.00764) | (0.00612) |
| Industry Return | | | | 0.218** | 0.267*** | 0.159 | 0.198** |
|  | | | | (0.0913) | (0.0801) | (0.129) | (0.0903) |
| Constant | 0.154*** | 0.0398 | 0.129*** | -0.642*** | -0.652*** | -0.298 | -0.147 |
|  | (0.0198) | (0.0312) | (0.0115) | (0.207) | (0.176) | (0.284) | (0.234) |
| R^2 | 0.108 | 0.029 | 0.211 | 0.238 | 0.209 | 0.275 | 0.351 |

First, let's look at the results of the three main variables that influence the IPO stock price return on the first day in [Table 8]. The stock price returns on the first day of the IPO are predicted well enough even when only one ASVI variable is used without controlling company, industry, or IPO-related characteristic variables. At this time, the coefficient value of ASVI has a significant positive value of 0.276 in the period from 2004 to 2008 and 0.321 in the period from 2009 to 2019. Both sections can be considered to have sufficiently high predictive power, but they have higher predictive power in the section after 2009. In other words, after 2009, when search data surged, SVI better measured the interest of individual investors, and thus the stock price return forecast on the first day of the IPO using this is better suited.

When only one media variable (Liu (2009)) is used without controlling company, industry, and IPO-related characteristic variables, as in ASVI, it has a significant positive value, so it predicts the stock price return on the first day of the IPO. However, when comparing the coefficient size, R^2, the predictive power is inferior to that of ASVI. And, unlike SVI, media variables have less predictive power in the period after 2009 than in the period before 2009. This is because, after 2009, as article data surged, there were many news that were related to actual investment or high-quality news as well as indiscriminately including the company name among the news that appeared when searched by company name.

When looking at the results of using only one price correction variable without controlling the company, industry, and IPO-related characteristic variables, the predictive power of the stock price return on the first day of the IPO is the strongest among the three main variables. Looking at the value of $R^2$, it is 23.5% for the interval before 2009 and 21.1% for the interval after 2009, which is the strongest predictor regardless of the interval. This is because the price correction ratio is the most direct measure of the demand for a specific IPO stock, and there is no ticker symbol for stocks before IPO, so it shows better predictive power than SVI using the company name as a search keyword.

Next, check the experimental results of adjusting the characteristic variables related to the company (asset size), industry (industrial return rate), and IPO (procurement size) to the three main variables that affect the stock price return on the first day of the IPO. After controlling the characteristic variables, the coefficient value of ASVI is 0.204 before 2009 and 0.279 after 2009, which is slightly lower than before controlling the characteristic variables, but still has a significant positive value. After controlling for characteristic variables, the coefficient values of the media variable and the price correction variable were also lowered in the period after 2009 than before 2009, but they still have a significant positive value. However, if all three main variables are included and characteristic variables are controlled, the predictive power of media variables loses significance. In the end, the only investor interest that can predict the share price return on the first day of the IPO remains SVI as an alternative value, and it becomes the only way to measure investor interest before the final public offering price is released. Of course, although the price correction variable's predictive power of stock price returns on the first day of the IPO is still the highest, this is a value that can be obtained after the public offering price is confirmed, and there is a problem that it is not possible to observe the trend of changes in interest.

Next, compare the results according to the sampling interval. It can be seen that the predictive power of ASVI increases markedly in the period after 2009 than in the period before 2009. This can be seen as a better measure of investor interest as investors search the Internet for investment information with the advent of mobile devices and data increase since 2009. In the remaining experimental results, it can be seen that the $R^2$ of the period after 2009 than in the period before 2009 appears lower in the case of the media variable and the price correction variable experiment, but rises in the case of ASVI. In other words, it can be seen that the predictive power of the stock price return on the first day of the IPO of the media variable and the price correction variable has recently decreased in the market, but the explanatory power of ASVI has increased. In the end, it can be said that the use of SVI in the recent market has increased even more than in the past market. Among the characteristic variables, in the case of procurement size, which is a variable related to IPO, it can be seen that the coefficient value is generally lowered after 2009 than before. This increases the speed of information transmission to investors and It seems that this is because the perception of investors has grown more than in the past.

## Chapter 6. Conclusion

Data such as excess transaction volume, excess returns, and news data, which previously measured investor interest, are not only a method of indirectly measuring investor interest, but also have various problems such as needing transaction data or selecting high-quality news. Existed. Today, due to the proliferation of mobile devices and the popularization of Internet use, the accessibility of search services has increased, and as a result, search data has rapidly increased. In addition, most individual investors who are not professional or institutional investors who have access to specialized terminals are forced to use a lot of search to obtain investment-related data. In this study, we investigated whether it is possible to directly measure the interest of individual investors using search data and whether the measured interest is still significant.

First of all, it was confirmed that the investor interest measurement method using SVI has a positive correlation with the existing interest measurement alternative values, but the degree is low. In addition, testing of both the pre-2009 and post-2009 sampling intervals on Russell 3000 stocks showed that SVI measures investor interest more quickly and accurately. In addition, it was confirmed that the investor interest measurement method using SVI has a greater explanatory power after 2009, when search data rapidly increased.

As a result of verifying the pressure of rising prices due to increased interest (Barber and Odean (2008)), including all the latest markets using SVI values, the increase in SVI still induces a real short-term price increase in Russell 3000 stocks. It has been shown that stock prices are returning. It also confirmed that even for IPO stocks in the latest markets, high SVI contributes to high stock returns on day one and regresses in the long run as well.

Search data is an objective indicator that can timely measure the interest of investors, and at the same time, it can be seen as the value that can best measure the interest of individual investors who do not need transaction data and do not need to select data separately. It is also the only value that can measure investor interest in IPO stocks.

As the mobile device became more popular, it was confirmed through various experiments that as search data increases and information delivery speed increases, investor interest measurement using search data has higher explanatory power and can be more useful than existing studies. In addition, it was confirmed that there is still a price pressure phenomenon caused by investor interest. In the future, various additional experiments can be expected by using the investor interest level using this increased search data.